\documentclass[12pt]{article}
\usepackage{amsmath}
\usepackage{graphicx}
\usepackage{amssymb}
\usepackage{amsfonts}
\usepackage{latexsym}
\def\be{\begin{equation}} \def\ee{\end{equation}}
\def\ba{\begin{eqnarray}} \def\ea{\end{eqnarray}} \def\part{\partial}
\def\nn{\nonumber}\def\lb{\label}

\begin{document}

\begin{center}
\begin{flushright}\begin{small}    LAPTH-1303/09
\end{small} \end{flushright} \vspace{1.5cm}
\huge{Phantom Black Holes} \huge{in Einstein-Maxwell-Dilaton Theory}
\end{center}

\begin{center}
{\small \bf G\'erard Cl\'ement $^{(a)}$}\footnote{E-mail address:
gclement@lapp.in2p3.fr}, {\small \bf Julio C. Fabris
$^{(b,c)}$}\footnote{E-mail address: fabris@pq.cnpq.br} \\
\ and
{\small \bf Manuel E. Rodrigues $^{(a,b)}$}\footnote{E-mail
address: esialg@gmail.com} \vskip 4mm

(a) \ LAPTH, Laboratoire d'Annecy-le-Vieux de Physique Th\'{e}orique
\\9,
Chemin de Bellevue - BP 110\\
74941 Annecy-le-Vieux CEDEX, France

\vskip 2mm (b) \ Universidade Federal do Esp\'{\i}rito Santo \\
Centro de Ci\^{e}ncias
Exatas - Departamento de F\'{\i}sica\\
Av. Fernando Ferrari s/n - Campus de Goiabeiras\\ CEP29075-910 -
Vit\'{o}ria/ES, Brazil \vskip 2mm

(c) \ IAP, Institut d'Astrophysique de Paris \\ 98bis, Bd Arago -
75014 Paris, France

\end{center}

\begin{center}
                                       Abstract
\end{center}
 We obtain the general static, spherically symmetric
solution for the Einstein-Maxwell-dilaton system in four dimensions
with a phantom coupling for the dilaton and/or the Maxwell field.
This leads to new classes of black hole solutions, with single or
multiple horizons. Using the geodesic equations, we analyse the
corresponding Penrose diagrams revealing, in some cases, new causal
structures.
\newpage

\section{Introduction}

Effective gravity actions emerging from string and Kaluza-Klein
theories contain a rich structure where, beside the usual
Einstein-Hilbert term, there is a scalar field, generically called a
dilaton, coupled to an electromagnetic field. The asymptotically
flat static black hole solutions for this Einstein-Maxwell-dilaton
(EMD) system \cite{gibbonsa,strom} differ from the usual
Reissner-Nordstr\"om solution of Einstein-Maxwell theory in that the
inner horizon is singular for a non-vanishing dilaton coupling.
Non-asymptotic flat static black hole solutions have also been
obtained in \cite{CHM} and further studied in
\cite{cedric,cedric1,glauber}. Brane configurations, leading also to
black hole solutions, have been largely studied in reference
\cite{grojean,cedric2} using essentially the EMD theory.

The aim of the present work is to study the structures of the black
holes of the EMD theory when a phantom coupling is considered. This
is done by allowing the scalar field or the Maxwell field (or both)
to have the ``wrong" sign \cite{gibbons}. The importance of such
extension of the normal (non-phantom) EMD theory is twofold. First,
from the theoretical point of view, string theories admit ghost
condensation, leading to phantom-type fields \cite{veneziano}. In
principle, a phantom may lead to instability, mainly at the quantum
level. But there are claims that these instabilities can be avoided
\cite{piazza}. The second motivation comes from the results of the
observational programs of the evolution of the Universe, specially
the magnitude-versus-redshift relation for the supernovae type Ia,
and the anisotropy spectrum of the cosmic microwave background
radiation: both observational programs suggest that the universe
today must be dominated by an exotic fluid with negative pressure.
Moreover, there is some evidence that this fluid can be phantom
\cite{han1,wmap}.

When the scalar field and/or the Maxwell field are allowed to
contribute negatively to the total energy, the energy conditions
(and specially the null energy condition $\rho + p \geq 0$) can be
violated, and the appearance of some new structures can be expected.
This is the case of wormhole solutions to Einstein-scalar
\cite{ellis} or Einstein-Maxwell-scalar \cite{kirill5} theory with a
phantom scalar field. Another instance is that of black hole
solutions to Einstein theory minimally coupled to a free scalar
field, which are forbidden by the no-hair theorem, but become
possible if the kinetic term of the scalar field has the wrong sign
(if the scalar field is phantom), as found both in 2+1 \cite{cold1}
and in 3+1 dimensions \cite{manuel1}. These phantom black holes have
all the characteristics of the so-called cold black holes
\cite{kirill2,kirill3}: a degenerate horizon (implying a zero
Hawking temperature) and an infinite horizon surface. Indeed, these
cold black holes appeared in the context of scalar-tensor theory
(e.g., in Brans-Dicke theory) in such circumstances that, after
re-expressing the action in the Einstein frame, the scalar field
comes out to be phantom.

The non-trivial dilatonic coupling of the scalar field with the
electromagnetic term adds new classes of black hole solutions. In
references \cite{gibbons,gao} some investigations on phantom black
holes in the context of EMD theory have been made, revealing some
interesting new species of black holes. For example, in the case of
a self-interacting scalar field in four dimensions, a phantom field
may lead to a completely regular spacetime where the horizon hides
an expanding, singularity-free universe \cite{kirill1}. Our goal
here is to obtain the most general static black hole solutions when
a phantom coupling is allowed for both the scalar and
electromagnetic fields of the EMD theory.

We will classify the different possible black hole solutions coming
out from the EMD theory for a phantom coupling of either the dilaton
field, of the Maxwell field, or of both. It is remarkable that many
of these new black holes have a degenerate horizon, hence a zero
Hawking temperature. We will also analyse the causal structure of
these black hole spacetimes. In some cases, the causal structure is
highly unusual, such that no two-dimensional Penrose diagram can be
constructed. Another possibility is that of a spacetime with an
infinite series of regular horizons separating successive
non-isometric regions. Geodesically complete black hole spacetimes
are also obtained.

This paper is organized as follows. In the next section we derive,
following the procedure of \cite{gerard1}, the general static
spherically symmetric solutions (phantom and non-phantom) of the EMD
theory. In section 3 the new black hole solutions are described in
detail. The Penrose diagrams of these new solutions are constructed
in section 4. In section 5 we present our conclusions.


\section{General solution}
\setcounter{equation}{0} Let us consider the following action: \be
S=\int dx^{4}\sqrt{-g}\left[  \mathcal{R}-2\,\eta_{1}
g^{\mu\nu}\nabla_{\mu}\varphi\nabla_{\nu }\varphi+\eta_{2} \,e^{
2\lambda\varphi}F^{\mu\nu}F_{\mu\nu}\right]  \label{action1}\; ,
\end{equation}
which is the sum of the usual Einstein-Hilbert gravitational term,
a dilaton field kinetic term, and a term coupling the Maxwell Lagrangian
density  to the dilaton, with the coupling constant $\lambda$ real.
The dilaton-gravity coupling constant $\eta_1$ can take either the value
$\eta_{1}=1$ ({\it dilaton}) or $\eta_{1}=-1$ ({\it anti-dilaton}). The
Maxwell-gravity coupling constant $\eta_{2}$ can take either the value
$\eta_{2}=1$ ({\it Maxwell}) or $\eta_{2}=-1$ ({\it anti-Maxwell}).
This action leads to the following field equations:
\begin{eqnarray}
\nabla_\mu\left[  e^{2\lambda\varphi}
F^{\mu\alpha}\right]&=&0\label{fe1}\; ,\\
\Box\varphi &=& - \frac{1}{2}\eta_{1}\eta_{2}\lambda e^{2\lambda\varphi}F^2
\; ,\label{fe2}\\
R_{\mu\nu}&=&2\eta_{1}\nabla_{\mu}\varphi\nabla_{\nu}\varphi+2\eta_{2}\,
e^{2\lambda\varphi}\left(\frac {1}{4}g_{\mu\nu}F^{2}
-F_{\mu}^{\;\;\sigma}F_{\nu\sigma}\right)\label{fe3}\, .
\end{eqnarray}

Let us write the static, spherically symmetric metric as
\begin{equation}
dS^{2}=e^{2\gamma(u)}dt^{2}-e^{2\alpha(u)}du^{2}-e^{2\beta(u)}d\Omega^{2}
\label{metric}\; .
\end{equation}

The metric function $\alpha$ can be changed at will by redefining
the radial coordinate. In the following we will assume the harmonic
coordinate condition
\be \alpha = 2\beta + \gamma \,.
\ee

We will also assume the Maxwell field to be purely electric (the purely
magnetic case may be obtained from this by the electric-magnetic
duality transformation $\varphi\rightarrow -\varphi$, $F \to
e^{-2\lambda\varphi}*F$). Integrating (\ref{fe1}), we obtain
\begin{equation}
F ^{10}(u)=q e^ {-2(\lambda\varphi+2\beta+\gamma)} \qquad
(F ^{2}=-2q^{2}e^{-4\beta-4\lambda\varphi}) \label{s1}\; ,
\end{equation}
with $q$ a real integration constant. Replacing (\ref{s1}) into equations, we
obtain the second order equations
\begin{eqnarray}
\varphi''  &=& -\eta_{1}\eta_{2}\lambda q ^{2}e^{2\omega} \label{e2bis}\; ,\\
\gamma'' &=&\eta_{2}q ^{2}e^{2\omega} \label{e1bis}\; ,\\
\beta'' &=&e^{2J}-\eta_{2}q ^{2}e^{2\omega} \label{e3bis}\; ,
\end{eqnarray}
with
\be
\omega=\gamma-\lambda\varphi\,, \quad J=\gamma+\beta \,,
\ee
and the constraint equation
\be\label{cons}
\beta^{'2} + 2\beta'\gamma' - \eta_1\varphi'^2 = e^{2J} - \eta_2q^2
e^{2\omega}\,.
\ee

By taking linear combinations of the equations (\ref{e2bis})-(\ref{e3bis}),
this system can be partially integrated to
\begin{eqnarray}
\varphi (u)&=&-\eta_{1}\lambda\gamma (u) +\varphi_{1}u+\varphi_{0}\; ,
\label{e1tertio}\\
\omega'^2 -Qe^{2\omega}&=& a^2\label{e2tertio}\; ,\\
J'^{2}-e^{2J}&=&b^{2}\label{e3tertio}\; ,
\end{eqnarray}
where
\be
\lambda_{\pm}=(1\pm\eta_{1}\lambda^{2})\,, \quad Q=\eta_{2}\lambda_{+}q^2\,,
\ee
and the integration constants $\varphi_{0},\varphi_{1}\in\mathbb{R}$,
$a,b\in\mathbb{C}$.

The general solution of (\ref{e2tertio}) is:
\begin{eqnarray}\label{om}
\omega(u)=\left\{\begin{array}{lr}
-\ln \left| \sqrt{\left| Q\right|} a^{-1}\cosh [a(u-u_{0})]\right| \quad
(a\in \mathbb{R}^{+}\; , \;Q\in\mathbb{R}^{-})\; , \\
a(u-u_{0}) \qquad (a\in\mathbb{R}^{+}\;, \; Q=0 )\; ,\\
-\ln \left| \sqrt{Q} a^{-1}\sinh [a(u-u_{0})]\right| \quad (a\in \mathbb{R}^+
\;,  \;Q\in\mathbb{R}^{+})\; ,\\
-\ln \left| \sqrt{Q} (u-u_{0})\right| \quad (a=0,\; Q\in\mathbb{R}^{+})\;,\\
-\ln \left| \sqrt{Q} \bar{a}^{-1}\sin [\bar{a}(u-u_{0})]\right| \quad (a
=i\bar{a},\; \bar{a},Q\in\mathbb{R}^{+})
\end{array}\right.
\end{eqnarray}
($u_0$ real constant). The general solution of (\ref{e3tertio}) is:
\begin{equation}\label{J}
J(u)=\left\{\begin{array}{lr}
-\ln \left|  b^{-1}\sinh [b(u-u_{1})]\right| \quad (b\in\mathbb{R}^{+})\;,\\
-\ln \left| u-u_{1}\right| \quad (b=0)\; ,\\
-\ln \left| \bar{b}^{-1}\sin [\bar{b}(u-u_{1})]\right| \quad (b=i\bar{b} ;\;
\bar{b}\in\mathbb{R}^{+})
\end{array}\right.
\end{equation}
($u_1$ real constant).

In this way, we have for $\lambda _{+} \neq 0$ the general solution of the
theory given by the action (\ref{action1}):
\begin{eqnarray}
\label{gs} \left\{\begin{array}{lr}
dS^{2}=e^{2\gamma}dt^{2}-e^{2\alpha}du^{2}-e^{2\beta}d\Omega^{2}\; ,\\
\alpha (u)=2J(u)-\gamma (u)\; ,\\
\beta (u)=J(u)-\gamma (u)\; , \\
\gamma (u)=\lambda _{+}^{-1}(\omega(u)+\lambda\varphi_{1}u+\lambda\varphi_0)
\; ,\\
\varphi (u)=\lambda _{+}^{-1}(-\eta_{1}\lambda\omega(u)+\varphi _{1}u+
\varphi _{0})\; ,\\
F=-q\; e^{2\omega (u)}du\wedge dt\;,
\end{array}\right.
\end{eqnarray}
where the integration constants should be related by the constraint
(\ref{cons}). We will analyze the case $\lambda _{+}= 0$ ($\eta_{1}=-1,
\lambda^2=1$), for which the function $\omega (u)$ is linear,
in subsection 3.2.

In the limit $u \to u_1$, the function $J(u)$ goes to $+\infty$,
corresponding from (\ref{J}) to spacelike infinity. So there are a priori two
disjoint solution sectors $u-u_1 > 0$ and $u-u_1 < 0$. However the ansatz
(\ref{metric}) is form-invariant under the symmetry $u-u_1 \to -(u-u_1)$,
which allows us to select e.g. the solution sector
\begin{equation}\lb{uneg}
u < u_1\,.
\end{equation}

Invariance of the metric ansatz under translations of the radial
coordinate $u$ also enables us to fix e.g. the integration constant
\be
u_{1}=0\,.
\ee

The solution (\ref{gs}) then depends on
the 6 parameters ($ q, a, b, u_{0}, \varphi _{0}, \varphi _{1}$)
which are related by the constraint equation following from
(\ref{cons}),
\begin{equation}
\lambda _{+}b^{2}=a^{2}+\eta _{1}\varphi
_{1}^{2}\label{constraint}\; .
\end{equation}

Moreover, two of the parameters may be fixed by imposing that at
infinity the space-time is Minkowskian and the dilaton field
vanishes. Hence, we end up with three independent parameters. Later,
imposing the analyticity of the solution across the horizon, the
number of free parameters will be reduced to only two.

All solutions written above with $\eta_1$ and $\eta_2$ positive (the
non-phantom case) have already been determined previously (see for example
\cite{gerard1} and references therein). To our knowledge, only the phantom
solutions with non-degenerate horizon have already been determined.


\section{New black hole solutions}
\setcounter{equation}{0}

From (\ref{om}) and (\ref{J}), with $u < 0$ according to
(\ref{uneg}), we have 15 different solutions which combine to form
the solution (\ref{gs}). The first and second solutions (\ref{om})
are necessarily phantom ($\eta_{1}=-1$ and/or $\eta_{2}=-1$). The
other ones can be normal ($\eta_{1}=\eta_{2}=1$), or phantom,
($\eta_{1}=\eta_{2}=-1,\lambda^{2}>1$ or $\eta_{1}=-\eta_{2}=-1,
\lambda^{2}<1$). As for the function $J(u)$, only the first solution
(\ref{J}) occurs in the normal case. This is because if $b=
i\bar{b}$ ($\bar{b}$ real)\footnote{This includes the case $b=0$.}
the constraint (\ref{constraint}) can be written \be a^2 + \bar{b}^2
= -\eta_1(\varphi_1^2 + \lambda^2\bar{b}^2)\,, \ee implying $\eta_1
= -1$ (anti-dilaton case). In this section we will discuss the new
black hole solutions, classified according to the type of the
solution (\ref{om}) for $\omega(u)$.
\subsection{The $\cosh$ solution}

The metric function $e^{2\gamma}$ for the choice of the first
solution (\ref{om}) can vanish only for $u \to -\infty$,
corresponding to the event horizon. The third solution (\ref{J}) for
$J$ obviously leads to a metric which is defined only in finite
intervals (in between zeroes of the sine), so it cannot have a
horizon. As for the second solution (\ref{J}), it can be written \be
dS^2 = e^{2\gamma}dt^2 - e^{-2\gamma}\left(dr^2 +
r^2d\Omega^2\right) \quad (r = -1/u)\,, \ee showing that the horizon
$u \to -\infty$ ($r=0$) is actually singular. So we restrict
ourselves to the first solution (\ref{J}).

To study the near-horizon behavior of the metric, it is
useful to transform to a radial coordinate  proportional to the near-horizon
geodesic affine parameter. From the Lagrangian for geodesic motion
\be
\mathcal{L}=\frac{1}{2}g_{\mu\nu}\dot{x}^{\mu}\dot{x}^{\nu}\,,
\ee
we obtain the first integral for geodesic motion in the equatorial plane
\begin{equation}
e^{4J}\dot{u}^{2}=E^{2}-e^{2\gamma }[\epsilon +L^{2}e^{-2\beta }]\; ,\label{22}
\end{equation}
where $E$ is the momentum conjugate to time (energy), $L$ is the
momentum conjugate to the azimutal angle (angular momentum), and
$\epsilon =1$ for timelike geodesics, $ \epsilon =0 $ for null
geodesics and $ \epsilon =-1 $ for spacelike geodesics. On the event
horizon this equation becomes
\begin{equation}
\label{23}
ds \sim e^{2J}du \sim b^2e^{2bu}du \; .
\end{equation}

This suggests performing the coordinate transformation
\begin{equation}\label{24}
x =e^{2b(u-u_0)}\,,
\end{equation}
leading to the line element
\begin{equation}\label{27}
dS^{2}=\frac{cx ^{n}}{(1+x^{m})^{2/\lambda _+}}dt^{2}-\frac{4b^{2}x _{1}
x^{1-n}(1+x ^m)^{2/\lambda _+}}{c(x-x _{1})^{2}}\bigg[ \frac{x _{1}
dx^{2}}{x(x-x _{1})^{2}}+d\Omega ^{2}\bigg]\; ,
\end{equation}
where according to (\ref{uneg}) $x<x_1$, with $x_{1}=e^{-2bu_{0}}$,
$c>0$ is a constant which can be fixed so
that the metric is asymptotically Minkowskian, and
\be\label{mn}
m =\frac{a}{b}\,, \quad n=\frac{m(1+\lambda\bar\varphi_{1})}{\lambda_{+}}
\ee
($\bar{\varphi_1}=\varphi_1/a$). This spacetime is asymptotically
flat, with the spatial infinity at $x=x_1$
($u=0$) and the event horizon at $x=0$. It is clear that the metric
is analytic near the event horizon only if $m$ and $n$ are positive
integers \cite{kirill2,cold1}.

The definition of $n$ and the constraint equation (\ref{constraint}) may be
rewritten as
\begin{eqnarray}\label{29}
\left\{\begin{array}{cr}
1+\lambda\bar{\varphi}_{1}=\lambda_{+} \frac{n}{m}\; ,\\
1+\eta_{1}\bar{\varphi}_{1}^{2}=\frac{\lambda_{+}}{m^{2}}\; .
\end{array}\right.
\end{eqnarray}

The first equation relates the integration constant
$\bar{\varphi_1}$ to the black hole quantum numbers $n$ and $m$.
Eliminating this constant between the two relations (\ref{29}), we
obtain the relation
\begin{equation}\label{cond1}
m = n \pm \lambda\sqrt{\eta_1(1 - n^2)} \, .
\end{equation}

The implications of this relation depend on the value of the horizon
degeneracy degree $n$. If $n=1$ (non degenerate horizon), then
necessarily also $m=1$; this can occur for any real $\lambda$ (with
$\eta_1$ and $\eta_2$ such that $\eta_2\lambda_+ < 0$). But if
$n\geq 2$ (degenerate horizon), which is only possible in the
anti-dilaton case $\eta_1 =-1$, then relation (\ref{cond1}) gives
the dilaton coupling constant $\lambda$ in terms of the black hole
``quantum numbers" $m$ and $n$: \be\label{ladis} \lambda^2 =
\frac{(m-n)^2}{n^2-1}\,. \ee This is such that $\lambda^2 < 1$
($\eta_2 = -1$) if $1 \le m \le 2n-1$, and $\lambda^2 > 1$ ($\eta_2
= +1$) if $m \ge 2n$.

Conversely, for a given value of the model parameters in
(\ref{action1}), there are three possibilities. Either $\eta_1 = +1$
or $\eta_1 = -1$ but $\lambda$ does not belong to the discrete set
of values (\ref{ladis}), and the only black hole solution is
non-degenerate ($m=n=1$); or $\eta_1 = -1$ and $\lambda\neq0$
belongs to this discrete set (with the appropriate sign for
$\eta_2$), and we have two distinct black hole solutions, one
degenerate, the other non-degenerate. Finally there is the third
possibility $\lambda = 0$ with $\eta_1 = \eta_2 = -1$, leading to a
tower of degenerate black hole solutions with $m=n$ above the ground
non-degenerate black hole $m=n=1$. Let us note that this last case
corresponds to the (non-dilatonic) Einstein-anti-Maxwell theory with
an additional massless scalar field coupled repulsively to gravity.
Recall that in the normal case ($\eta_1 = +1$), the second equation
(\ref{29}) with $\lambda_+ = 1$ has the only solution $m=1$ with
$\bar{\varphi}_{1} =0$, leading to the Reissner-Nordstr\"{o}m black
hole with constant scalar field. This is the well-known no-hair
theorem, which is no longer true in the present phantom
(anti-scalar) case.

To put the cosh solution in a more familiar form, we use the coordinate
transformation \cite{gerard1},
\begin{equation}\label{30}
u=\frac{1}{(r_{+}-r_{-})}\ln \left( \frac{f_{+}}{f_{-}}\right)\; ,\qquad
f_{\pm}=1- \frac{r_{\pm}}{r}\; ,
\end{equation}
with
\begin{equation}\label{30.1}
r_\pm = \pm \frac{2a}{1 + e^{\mp 2au_{0}}} \qquad (r_{+}-r_{-} = 2a)\;.
\end{equation}
The coordinate $x$ defined by (\ref{24}) is related to $r$ by \be
x^m = -\frac{r_-f_+}{r_+f_-}\,. \ee

In the case of a non-degenerate
horizon ($n=m=1$), the solution takes the following form:
\begin{eqnarray}\label{32}
dS^{2}&=&f_{+}f_{-}^{\frac{\lambda _{-}}{\lambda _{+}}}dt^{2}
-f_{+}^{-1}f_{-}^{-\frac{\lambda _{-}}{\lambda _{+}}}dr^{2}
-r^{2}f_{-}^{1-\frac{\lambda _{-}}{\lambda _{+}}}d\Omega^{2} \;  , \nonumber\\
F&=&-\frac{q}{r^2}dr\wedge dt\; ,\;
e^{-2\lambda\varphi}=f_{-}^{1-\frac{\lambda_{-}}{\lambda_{+}}}
\; ,
\end{eqnarray}
where we have chosen the integration constants so that at spatial
infinity the metric is Minkowskian and the dilaton vanishes, i.e.
$\varphi_0 = \omega(0) = 0$ in (\ref{gs}). The corresponding black
hole mass $M$ and charge  $q$ are \be M=\frac{1}{2}\left(
r_{+}+\frac{\lambda_{-}}{\lambda_{+}}r_{-}\right)\; , \quad
q=\pm\sqrt{\frac{r_{+}r_{-}}{\eta_2\lambda_{+}}}\; . \ee

The solution (\ref{32}) has the same form as the normal black hole
solution of \cite{gibbonsa,strom}, the only difference being that
here we have $r_+ > 0$ but $r_- < 0$. It was previously obtained by
analytic continuation of the normal solution by Gibbons and Rasheed
\cite{gibbons} in the cases ($\eta_2 = -1, \eta_1 = +1$) and $\eta_1
= -1$ with $\eta_2 = +1$ (for $a^2 > 1$) or $\eta_2 = -1$ (for $a^2
< 1$), and was later generalized to higher-dimensional black holes
(in the anti-dilaton case $\eta_1 = -1$ with $\eta_2 = +1$) by Gao
and Zhang \cite{gao}, and to higher-dimensional black branes by Grojean
et al \cite{grojean}.

For the case of a degenerate horizon ($\eta_1=-1$), performing the
transformation (\ref{30}), we obtain the form (valid only outside the
horizon, $r > r_+$)
\begin{eqnarray}\label{34}
dS^{2} &=& g_{+}^{n}g_{-}^{\nu-n}dt^{2}
 - \frac{(r_{+}-r_{-})^{2}}{m^{2}}\frac{g_{+}^{1-n}g_{-}^{1+n-\nu}}
 {(g_{+}-g_{-})^{2}}\times\nonumber\\
 &&\times\biggr[ \frac{(r_{+}-r_{-})^{2}
 (g_{+}g_{-})^{1-2m}}{m^{2}r^{4}(g_{+}-g_{-})^{2}}dr^{2}+d\Omega^{2}
 \biggl]\,, \\
 F&=&-\frac{q}{r^2}dr\wedge dt\,,\quad
e^{-2\lambda\varphi}=g_{+}^{m-n}g_{-}^{m+n-\nu} \; ,
\end{eqnarray}
where $g_{\pm} = f_{\pm}^{1/m}$, and 
\be 
\nu = \frac{2m}{\lambda_+}= \frac{2m(1-n^2)}{(m-n)^2+1-n^2}\,, 
\ee
with the mass and charge
\be
 M=\frac{1}{2m}\left[nr_{+}+(\nu-n)r_{-}\right]\,,\quad
q=\pm\sqrt{\frac{r_{+}r_{-}}{\eta_2\lambda_{+}}}\; .
\ee


\subsection{The linear solution}

In the special case $\lambda =\pm1$ and $\eta_{1}=-1$, the constant
$\lambda_+$ vanishes, and $\omega (u)$ is given by the second
solution (\ref{om}), which depends linearly on the $u$ coordinate.
In this case, the functions $\gamma(u)$ and $\varphi(u)$ in
(\ref{gs}) are replaced by
\begin{eqnarray}\label{35}
\gamma (u)&=&\gamma_{1}u+\gamma_{0}+ \frac{q_1}{2}e^{2a (u-u_{0})}\; ,\\
\varphi (u)&=& \pm[(\gamma_{1} - a)u+(\gamma_{0} + a u_{0})+
\frac{q_1}{2}e^{2a (u-u_{0})}]\,,
\end{eqnarray}
where $\gamma_{0}$ and $\gamma_{1}$ are integration constants, and
$q_{1}=\eta_{2}q^{2}/2a^{2}$. Choosing again the $\sinh$ solution in
(\ref{J}), and performing the coordinate transformation (\ref{24}),
we obtain the metric,
\begin{eqnarray}
\label{37}
dS^{2}&=&cx^{n}\exp{\left[
q_{1}x^{m}\right]}dt^{2}\nonumber\\
&-&\frac{4b^{2}x_{1}x^{1-n}}{c(x-x_{1})^{2}}
\exp{\left[-q_{1}x^{m}\right]}\left[
\frac{x_{1}dx^{2}}{x(x-x_{1})^{2}}+d\Omega^{2}\right]\; ,
\end{eqnarray}
where again $x_{1}=e^{-2bu_{0}}$, $c>0$, and
\be m=\frac{a}{b}\,, \quad
n=\frac{\gamma_{1}}{b}\,.
\ee

As in the case of the cosh solution, the spacetime is asymptotically
flat, with the spatial infinity at $x=x_1$ and the event horizon at
$x=0$. Again, for the metric to be analytic near the event horizon,
the constants $m$ and $n$ must be positive integers. These are not
independent, because of the constraint (\ref{constraint}) which now
reads
\be
b^2 -a(2\gamma_1-a) = 0 \quad \Rightarrow \quad m(2n-m) =
1\,.
\ee

This is just the relation (\ref{ladis}) for $\lambda^2 =
1$, and it is clear that its only solution in terms of integers is
$m=n=1$. Performing the transformations  $x=x_{1}f_{+}$, with
$f_{+}=1-r_{+}/{r}$, $r_{+}=2b$, we recover the linear solution
previously discussed by Gibbons and Rasheed \cite{gibbons} and by
Gao and Zhang \cite{gao}:
\begin{eqnarray}\label{37.1}
dS^{2}&=&f_{+}\exp{[q_1(f_{+}-1)]}dt^{2}-\exp{[-q_1(f_{+}-1)}
\left(\frac{dr^{2}}{f_{+}}+r^{2}d\Omega^{2}\right)\; ,\\
F&=&-\frac{q}{r^{2}}dr\wedge dt\,,\quad
e^{-2\lambda\varphi}=\exp[-q_1(f_{+}-1)]\,, \ea with \be
M=\frac{1+q_1}2r_{+}\,, \quad q=\pm r_+
\sqrt{\frac{\eta_{2}q_1}2}\,. \ee


\subsection{The phantom $\sinh$ solutions}

The third solution (\ref{om}) leads to a larger spectrum of phantom
black holes, with $\eta_1 < 0$ ($\eta_1 > 0$ with $Q > 0$ implies
$\eta_2 > 0$, leading to normal black holes). The event horizon can
correspond either to $u \to -\infty$, with only the first solution
(\ref{J}) for $J(u)$, or (if $\lambda_+ < 0$) to $u = u_0$, with the
three possibilities (\ref{J}). There is also in the first case the
possibility of a non-asymptotical flat black hole spacetime when the
singularity $u=u_0$ of the metric coincides with spacelike infinity
$u=0$ \cite{cedric2}.

We first consider the case where $J(u)$ is given by the first
expression of (\ref{J}). Performing as in the cosh case the
coordinate transformation (\ref{24}), we obtain the following line
element: \be\label{sinh} dS^{2}=\frac{cx ^{n}}{|1-x^{m}|^{2/\lambda
_+}}dt^{2}-\frac{4b^{2}x _{1} x^{1-n}|1-x ^m|^{2/\lambda _+}}{c(x-x
_{1})^{2}}\bigg[ \frac{x _{1} dx^{2}}{x(x-x _{1})^{2}}+d\Omega
^{2}\bigg]\; , \ee where again $x<x_1$ with $x_{1}=e^{-2bu_{0}}$,
and $c>0$. The real numbers $m$ and $n$, given by (\ref{mn}), are
again related by (\ref{cond1}). There are three possibilities
according to the relative values of $x_1$ and 1.

a) If $x_1 < 1$ ($u_0 > 0$), the event horizon is located at $x=0$. As in
the cosh case, this is regular if both $m$ and $n$ are integer. The discussion
on the possibility of degenerate horizons is identical to that following Eq.
(\ref{cond1}), provided that $\eta_2$ is replaced by $-\eta_2$
($\eta_2\lambda_+$ is positive in the sinh case). Performing the coordinate
transformation (\ref{30}), with
\begin{equation}\label{30.1}
r_\pm = \pm \frac{2a}{1 -e^{\mp 2au_{0}}} \qquad (r_{+}-r_{-} = 2a)\;,
\end{equation}
we recover in the case $m=n=1$ the metric (\ref{32}), with now $0<r_-<r_+$.
These solutions were previously obtained in \cite{gibbons}.
For the cases with degenerate horizon ($\eta_1=-1, n\geq 2$), we obtain
the functional form (\ref{34}).

b) In the intermediate case $x_1 = 1$ ($u_0 = 0$), the event horizon
is located, as in the first case, at $x = 0$, with $m$ and $n$
integer. However the resulting solution is no longer asymptotically
flat \cite{cedric2}. Performing the coordinate transformation
$x=f_+$ with $r_+ = 2bc^{-1/2}$ and rescaling the time coordinate,
(\ref{sinh}) becomes: \be\label{sinh1} dS^{2}=\frac{f_+
^{n}}{(1-f_+^{m})^{2/\lambda _+}}dt^{2}- \frac{(1-f_+^m)^{2/\lambda
_+}}{f_+^n}[dr^2 + r(r-r _+)d\Omega ^{2}]\,. \ee

This has the asymptotic behavior \be\lb{naf} dS^2 \sim
\bigg(\frac{r}{r_+}\bigg)^{2/\lambda_+}dt^2 -
\bigg(\frac{r}{r_+}\bigg)^{-2/\lambda_+}\bigg(dr^2 + r^2
d\Omega^2\bigg)\,, \quad (r \to \infty) \ee with $0 < \lambda_+ \le
1$ for $\eta_2 > 0$, and $\lambda_+ < 0$ for $\eta_2 < 0$.

c) If $x_1 > 1$ ($u_0  < 0$), then the spacetime is asymptotically flat, but
the event horizon is located at $x = 1$,
provided $\lambda_+ < 0$, implying both $\eta_1<0$ and $\eta_2<0$, together
with $\lambda^2 > 1$. This is regular if $\lambda_+ = -2/p$, i.e.
\be
\lambda^2 = \frac{p+2}p\,,
\ee
with $p$ positive integer (but $m$ and $n$ no longer necessarily integer).

This possibility also occurs when the second or third solution
(\ref{J}) are combined with the third solution (\ref{om}) with $u_0
< 0$. In these cases the coordinate transformation (\ref{24}) leads
to an unwieldy form of the metric. A more manageable expression for
the metric in these three cases with $u_0<0$ is
\be\lb{horu0} dS^2 =
ch^pdt^2 -c^{-1}h^{-p}e^{2J}(e^{2J}du^2 + d\Omega^2)\,,
\ee with
\ba
h(u) &=& e^{-\lambda\varphi_1u}\sinh a(u-u_0)\,, \\
e^{2J(u)} &=& \frac{b^2}{\sinh^2bu}\,, \;\; {\rm or} \;\; \frac1{u^2}\,,
\;\; {\rm or} \;\; \frac{\bar{b}^2}{\sin^2\bar{b}u}\,.
\ea

\subsection{The $a=0$ solutions}

If $u_0 \ge 0$, the event horizon is again at $u \to -\infty$
provided $J(u)$ is given by the first or second solution (\ref{J}).
In the first case ($b^2>0$), the coordinate transformation
(\ref{24}) puts the metric in a form similar to (\ref{sinh}), with
$n = (\lambda/\lambda_+)(\varphi_1/b)$ and $m \to 0$, so that $x^m$
is replaced by a logarithm, which is clearly non-analytic. So this
possibility does not lead to a regular black hole. In the case $a =
b = 0$ (implying $\varphi_1=0$), the coordinate transformation $u =
-1/r$ leads to the metric \be\lb{ab0} dS^2 =
\bigg(\frac{u_0r}{1+u_0r}\bigg)^{2/\lambda_+}dt^2 -
\bigg(\frac{u_0r}{1+u_0r}\bigg)^{-2/\lambda_+}(dr^2 +
r^2d\Omega^2)\,, \ee which is analytic if $\lambda_+ = 2/(p+1)$ ($p$
positive integer), implying $\eta_2 > 0$, so that the phantom case
corresponds to $\eta_1 < 0$, and \be \lambda^2 = \frac{p-1}{p+1} \,.
\ee

Note that this is a special case of relation (\ref{ladis}) for
$m = 1$, $n = p$.

If on the other hand $u_0 < 0$, the event horizon is at $u=u_0$ and is regular
if $\lambda_+ = -2/p$ with $p$ a positive integer. Again, this implies
$\eta_1 < 0$, $\eta_2 < 0$, and $\lambda^2 > 1$. The constraint
(\ref{constraint}) reduces in this case to $b^2 = p\varphi_1^2/2 \ge 0$.
The resulting metric is of the form (\ref{horu0}) with
\be\lb{ha0}
h(u) = e^{-\lambda\varphi_1u}(u-u_0)\,.
\ee

For $b = 0$ ($\varphi_1=0$), the coordinate transformation $u = -1/r$
leads to the particularly simple form of the metric
\be\lb{ab0m}
dS^2 = f_+^pdt^2 - f_+^{-p}(dr^2 + r^2d\Omega^2)\,,
\ee
with $r_+ \equiv -1/u_0$.

\subsection{The $\sin$ solution}

In this case the metric is only defined in finite intervals and so
cannot have a horizon at $u \to -\infty$. Again, the event horizon
can only be at $u=u_0<0$, and is regular if $\lambda_+ = -2/p$ ($p$
positive integer). The constraint (\ref{constraint}) becomes in this
case \be \frac{2b^2}p = \bar{a}^2 + \varphi_1^2\,, \ee so that
necessarily $b^2 > 0$, corresponding to the first solution (\ref{J})
for $J(u)$. The resulting metric can be put in the form
(\ref{horu0}), with \ba
h(u) &=& e^{-\lambda\varphi_1u}|\sin \bar{a}(u-u_0)|\,, \lb{hsin}\\
e^{2J(u)} &=& \frac{b^2}{\sinh^2bu}\lb{Jsinh}\,.
\ea

\section{Geodesics and the Penrose diagrams}
\setcounter{equation}{0}

\subsection{The $\cosh$ solution}

The global structure of the new black hole spacetimes may be
determined by analyzing the geodesic equation (\ref{22}), written in
terms of the $x$ coordinate of (\ref{27}),
\begin{equation}
\frac{4b^{2}x_{1}^{2}\dot{x}^{2}}{(x-x_{1})^4}
=E^{2}-\frac{cx^n}{(1+x^m)^{2/\lambda_{+}}}\left[ \epsilon +
\frac{L^{2}cx^{n-1}(x-x_{1})^2}{4b^{2}x_{1}
(1+x^m)^{2/\lambda_{+}}}\right]\,,\label{geocosh}
\end{equation}
together with the conformal form of the metric (\ref{27}),
\begin{equation}\label{38}
dS^{2}=H(x)\left[ dt^{2}-dy^{2}-F(x)d\Omega ^{2}\right]\; ,
\end{equation}
with
\begin{eqnarray}\label{39}
dy&=&\pm
\frac{2bx_{1}}{c}\,\frac{x^{-n}(1+x^{m})^{2/\lambda_{+}}}
{(x-x_{1})^{2}}dx\; ,\\
H(x)&=&\frac{cx^{n}}{(1+x^{m})^{2/\lambda_{+}}}\; , \\
F(x)&=&
\frac{4b^2x_1}{c^2}\frac{x^{1-2n}(1+x^m)^{4/\lambda_+}}{(x-x_{1})^2}\;.
\end{eqnarray}

The various limits which should be analysed are $x\rightarrow x_{1}$
(the spatial infinity), $x\rightarrow 0$ (the event horizon), as well as
possible coordinate singularities at $x\rightarrow -1$ and
$x\rightarrow -\infty$.
In the limit $x\rightarrow x_{1}$, we obtain $y\rightarrow
\pm\infty\;$, with $F(x)=y^2$ and $H=$cst, showing that the metric
(\ref{38}) is asymptotically Minkowskian.
In the limit $x\rightarrow 0$, on the other hand, we obtain $y\rightarrow
\pm\infty\;$ and $H\rightarrow 0$. This characterizes a horizon, in the
present case the event horizon of the black hole which, as previously
discussed, is regular for all integer values of $n$ and $m$.

The analysis of the other possible coordinate singularities, at
$x\rightarrow -1$ or $x \rightarrow - \infty$ depends on the parity
of $m$. Let us summarize the different cases.
\begin{enumerate}
\item $m$ even. In this case, the metric (\ref{38}) is regular at
$x = -1$, which corresponds to a finite value of $y$. The nature of
the coordinate singularity at $x \rightarrow -\infty$ depends on the
value of $\lambda_+$.
\begin{enumerate}
\item If $\lambda_+ < 0$, ($\lambda^2>1$, which implies from (\ref{cond1})
$n < (m^2+1)/2m$, i.e. $n \leq m/2$) then $H \rightarrow \infty$ and
$y \rightarrow 0$. Hence, this is a center, a singularity where the
geodesics stop. If $n$ is odd the Penrose diagram is similar to the
Schwarzschild diagram (Fig. 1), since there is an inversion of the
light cone due to the change of signature ($+---$) $\to$ ($-+--$).
If $n$ is even, the two-dimensional light cone remains unchanged, so that
the Penrose diagram is similar to the extreme Reissner-Nordstr\"om diagram
(Fig. 2). However the full four-dimensional signature does change when the
horizon is crossed, ($+---$) $\to$ ($+-++$), so that geodesic motion in
this spacetime should differ significantly from that in extreme
Reissner-Nordstr\"om spacetime (see the discussion in \cite{kirill2}).
\item If $\lambda_+ > 0$ ($n > m/2$), we note that Eq. (\ref{cond1}) implies
$$\lambda_+ = 2m/(n + 1) - (m-1)^2/(n^2-1) < 2m/(n + 1),$$
so that $y \rightarrow \infty$ with $H \to 0$, corresponding to a
horizon. To analyse whether geodesics can be continued through this
horizon, we rewrite the geodesic equation (\ref{geocosh}) in terms
of the transformed coordinate $z = -1/x$,
\begin{eqnarray}
\frac{4b^{2}z_{1}^{2}\dot{z}^{2}}{(z-z_{1})^4}
&=&E^{2}-\frac{c(-1)^nz^{-n+2m/\lambda_+}}{(1+z^m)^{2/\lambda_{+}}}\times\nonumber\\
&&\times\bigg[ \epsilon
+\frac{L^{2}c(-1)^nz^{-n-1+2m/\lambda_+}(z-z_{1})^2}{4b^{2}z_{1}
(1+z^m)^{2/\lambda_{+}}}\bigg]  \label{geocoshi}
\end{eqnarray}
($z_1 = -1/x_1$). This is not analytic at $z=0$, where geodesics
terminate (singular horizon), unless \be\lb{horinf1} 2m/\lambda_+ =
n + p\,, \ee with $p$ a positive integer. In view of the definition
of $\lambda_+$ and of (\ref{cond1}), this implies the equation
\be\lb{horinf2} (n+p)(m-n)^2 = (n+p-2m)(n^2-1)\,, \ee which can be
solved in terms of integers only if either \ba\lb{mnp}
& \alpha) & m=n=1 \qquad (p\;\; {\rm arbitrary})\,, \nn \\
& \beta) & m = p = 1 \qquad (n\;\; {\rm arbitrary})\,. \\
& \gamma) & m = n = p \qquad (\lambda=0)\,, \nn
\ea (the first two
possibilities are excluded for $m$ even). So generically this case
corresponds to a null singularity, leading to the diagram of Fig. 3
if $n$ is odd, or Fig. 4 (with the signature ($+-++$) in region II)
if $n$ is even.
\item In the case $\lambda_+ =1$ with $m = n= p$ even integer
(Einstein-anti-Maxwell-anti-scalar case $\lambda = 0$, $\eta_1 =
\eta_2 = - 1$), $x \rightarrow -\infty$ is a regular horizon. The
metric (\ref{27}) is in this case
\begin{equation}
dS^{2}=\frac{cx ^{n}}{(1+x^{n})^{2}}dt^{2}-\frac{4b^{2}x _{1}
x^{1-n}(1+x ^n)^{2}}{c(x-x _{1})^{2}}\bigg[ \frac{x _{1}
dx^{2}}{x(x-x _{1})^{2}}+d\Omega ^{2}\bigg]\,.
\end{equation}
This is form-invariant under the combined inversion $x \to z = 1/x$,
$x_1 \to z_1 = 1/x_1$, which transforms into each other the two
horizons $x = 0$, $z = 0$ and the two spacelike infinities $x =
x_1$, $z = z_1$. This structure is illustrated in the diagram of
Fig. 5, which differs from the Penrose diagram for Kerr spacetime in
that the two horizons are evenly degenerate, and the signature in
region II is ($+-++$). This spacetime is geodesically complete.
\end{enumerate}
\item $m$ odd. In this case, there is a coordinate singularity at $x = -1$.
Near this region, putting $x = - 1 + z$ with $z \to 0$, we find $H
\sim z^{-2/\lambda_+}$ and $y \sim z^{2/\lambda_+ + 1}$, with the
geodesic equation
\begin{equation}\lb{geo-1}
\frac{4b^2x_1^2\dot{z}^2}{(1+x_1)^4} \simeq E^2 -
c'z^{-2/\lambda_+}\left[ \epsilon -
\frac{L^2c'(1+x_1)^2}{4b^{2}x_1}z^{-2/\lambda_+}\right]
\end{equation}
($c'=(-1)^ncm^{-2/\lambda_+}$). Now we find the following
structures:
\begin{enumerate}
\item For $\lambda_+ < -2$, $H \rightarrow 0$ and $y \rightarrow 0$, so that
$x = - 1$ corresponds to a singularity, and the Penrose diagram is
that of Schwarzschild (Fig. 1) for $n$ odd and that of extreme
Reissner-Nordstrom (Fig. 2) for $n$ even.
\item For $-2 \le \lambda_+ < 0$, $H \rightarrow 0$ and $y \rightarrow
\infty$, so that $x = - 1$ corresponds to a horizon. However, unless
$-2/\lambda_+ = p$, with $p$ a positive integer, this horizon is
singular (null singularity). The corresponding Penrose diagram is
represented by Fig. 3 if $n$ is odd, and by Fig. 4 if $n$ is even.
\item The case $\lambda_+ = -2/p$ with $p$ integer ($\eta_1=-1$, $\eta_2=+1$,
$\lambda^2 = (p+2)/p$) leads, from (\ref{cond1}), to the equation
\be\lb{l2p} p(m-n)^2 = (p+2)(n^2-1)\,,
\ee
which can be solved in
terms of integers in two subcases. A first solution is
\be\label{alpha} m=n=1\,,
\ee
and $p$ an arbitrary integer; in this
subcase the metric (\ref{32}) takes the simple analytic form
\be
dS^{2}=f_{+}f_{-}^{-(p+1)}dt^{2} - f_{+}^{-1}f_{-}^{p+1}dr^{2}
-r^{2}f_{-}^{p+2}d\Omega^{2} \,,
\ee
with the inner horizon at $r=0
> r_-$. The other solution is
\be\label{beta}
m=2n+1\,, \quad p=n-1\,.
\ee

In both subcases the geodesics can be continued until $x \rightarrow
- \infty$ ($r = r_-$). In this limit, we obtain $H \sim x^{n+mp} \to
\infty$ and $y \sim x^{-n-mp-1} \rightarrow 0$, corresponding to a
singularity. In the first subcase (\ref{alpha}), the maximally
extended spacetime is represented by a diagram (Fig. 6) similar to
that of Reissner-N\"ordstrom (but with signature ($+-++$)in region
III) if $p$ is odd, or by the diagram of Fig. 7 (further discussed
below) if $p$ is even. In the second subcase (\ref{beta}), the
spacetime is represented by the diagram of Fig. 7 if $n$ is odd ($p$
even), or the diagram of Fig. 8 (with signature ($+-++$) in region
II) if $n$ is even ($p$ odd).
\item For $\lambda_+ > 0$, $H \to \infty$ and $y \to 0$ so that $x = -1$
corresponds to a singularity. As in case (2a), we find the
Schwarzschild diagram (Fig. 1) for $n$ odd and the extreme
Reissner-Nordstrom diagram (Fig. 2) for $n$ even.
\end{enumerate}
\end{enumerate}

Note that the global structure of the black hole spacetimes with a
non-degenerate, or oddly degenerate, outer horizon $r=r_+$ (order
$n$), followed by an evenly degenerate inner horizon $r=0$ (order
$p$) hiding a spacelike singularity $r=r_-$ , cannot be represented
by a two-dimensional Penrose diagram. The diagram of Fig. 7
represents faithfully the global topology of these spacetimes, at
the price of representing the infinite sequence of bifurcate null
asymptotically Minkowskian boundaries as dotted vertical (timelike)
lines, each corresponding to the past and future null infinities of
two contiguous regions I.

\subsection{The linear solution}

For the metric (\ref{37}) with $m=n=1$, the geodesic equation
(\ref{22}) becomes
\begin{equation}\label{40}
\frac{4b^{2}x_{1}^{2}\dot{x}^{2}}{(x-x_{1})^4}
=E^{2}-cxe^{q_{1}x}\left[ \epsilon
+\frac{L^2c}{4b^{2}x_{1}}(x-x_{1})^{2}e^{q_{1}x}\right]\,,
\end{equation}
and the conformal form of this metric is (\ref{38}), with
\be\lb{41}
dy=\pm \frac{2bx_1}{c}\frac{e^{-q_1x}}{x(x-x_1)^2}
dx\,.
\ee

As before, the limit $x\rightarrow x_{1}$ corresponds to the asymptotically
Minkowskian region, and $x\rightarrow 0$ to the regular event horizon.
The analysis of the limit $x\rightarrow -\infty$ depends on the sign
of $q_1$.
\begin{enumerate}
\item For $q_ 1<0$ ($\eta_2=-1$), $y\rightarrow 0$,
so that $x =-\infty$ corresponds to a singularity, and the Penrose diagram
is that of Schwarzschild (Fig. 1).
\item For $q_ 1>0$ ($\eta_2=+1$), $y\rightarrow \pm\infty$, corresponding
to a horizon. Near this horizon the geodesic equation (\ref{40})
reads, in terms of the variable $z = -1/(x-x_1)$ ($z \to 0$)
\begin{equation}
4b^{2}x_{1}^{2}\dot{z}^{2}
\simeq E^{2}+c'z^{-1}e^{-q_{1}/z}\left[ \epsilon
+\frac{L^2c'}{4b^{2}x_{1}}z^{-2}e^{-q_{1}/z}\right]
\end{equation}
($c' = ce^{q_1x_1} > 0$). The effective potential goes to 0 for $z \to +0$,
but goes to $+\infty$ for $z \to -0$ and $L^2>0$ with $\epsilon > 0$ or
$L^2 > 0$, so that radial timelike geodesics and all non-radial geodesics
terminate due to an infinite potential barrier. So the apparent horizon
$x\rightarrow -\infty$ is actually a null singularity. The corresponding
Penrose diagram is represented by Fig. 4 (with in this case the signature
(- + - -) in region II).
\end{enumerate}

\subsection{The phantom $\sinh$ solutions}

As we have seen in Subsect. 3.3, there are three cases, according to
the value of $u_0$.
\begin{enumerate}
\item For $u_0 > 0$, $J(u)$ is necessarily given by the first solution
(\ref{J}), and the metric can be written in the form (\ref{sinh}), which
is asymptotically Minkowskian in the limit $x\rightarrow x_1 < 1$.
For $m$ and $n$ integer, the metric can be
extended across the event horizon (of order $n$) at $x = 0$. The next
possible coordinate singularity is at $x = -1$, where the analysis can be
carried over from the cosh case, provided the parity of $m$ is changed:
\begin{enumerate}
\item $m$ odd. In this case, $x=-1$ is regular, and the geodesics extend
until $x=-\infty$. The nature of this coordinate singularity depends
on the value of $\lambda_+$.
\begin{enumerate}
\item For $\lambda_+ < 0$, as in the cosh case, $x = -\infty$ is a singularity.
The Penrose diagram is given by Fig. 1 for $n$ odd, and Fig. 2 for $n$ even.
\item For a generic value of  $\lambda_+ > 0$, $x = -\infty$ is a
null singularity. The Penrose diagram is given by Fig. 3 for $n$
odd, and Fig. 4 for $n$ even.
\item For $m=n=1,\lambda_+ =2/(p+1)$ (the first solution of
(\ref{horinf1})-(\ref{horinf2})), with $p$ positive integer, the
metric (\ref{32}) reads
\be
dS^{2}=f_{+}f_{-}^pdt^{2} - f_{+}^{-1}f_{-}^{-p}dr^{2}
-r^{2}f_{-}^{-p+1}d\Omega^{2} \,.
\ee

This metric has two regular horizons
$r_+$ ($x=0$)and $r_-$ ($x = -\infty$) hiding the singularity $r=0$.
For $p$ odd, the Penrose diagram is of the "normal"
Reissner-Nordstr\"om type, with signature ($+---$) in region III
(Fig. 6). For $p$ even, the diagram is that of Fig. 7.
\item For $m=p=1,\lambda_+ =2/(n+1)$, the coordinate transformation
$x = 1/z < 0$ leads to the metric
\ba dS^2 &=&
\frac{(-1)^{n+1}cz}{(1-z)^{n+1}}dt^2 - \frac{(-1)^{n+1}4b^{2}z
_{1}(1-z)^{n+1}}{c(z-z _{1})^{2}}\times\nonumber\\
&&\times\bigg[ \frac{z _{1} dx^{2}}{z(z-z
_{1})^{2}}+d\Omega ^{2}\bigg]\,,
\ea
($z_1 = 1/x_1$). This can be
continued across the horizon $z = 0$ until the singularity $z=1$.
The spacetime is represented by the normal Reissner-Nordstr\"om
diagram (Fig. 6) if $n$ is odd, and by Fig. 8 if $n$ is even.
\item For $m=n=p$ ($\lambda_+=1$, Einstein-Maxwell-anti-scalar
case), the metric
\begin{equation}\lb{emas}
dS^{2}=\frac{cx ^{n}}{(1-x^{n})^{2}}dt^{2}-\frac{4b^{2}x _{1}
x^{1-n}(1-x ^n)^{2}}{c(x-x _{1})^{2}}\bigg[ \frac{x _{1}
dx^{2}}{x(x-x _{1})^{2}}+d\Omega ^{2}\bigg]\,,
\end{equation}
is form-invariant under the combined inversion $x \to z = 1/x$, $x_1
\to z_1 = 1/x_1$. The difference with the Einstein-Maxwell-scalar
case is that now to $x_1<1$ corresponds $z_1>1$. The coordinate $z$
increases from $z=-\infty$ (event horizon) through $z=0$ (second
horizon) to the singularity $z =1$. As $n$ is odd, the Penrose
diagram is again of the normal Reissner-Nordstr\"om type (Fig. 6).
\end{enumerate}
\item{$m$ even}. In this case, there is a coordinate singularity at
$x=-1$. Putting $x = -1 + z$, we obtain for the geodesic equation
near $z = 0$ the form (\ref{geo-1}). Therefore the analysis proceeds
as in case 2 of Subsect. 4.1, except for the case $\lambda_+ = -2/p$
which does not occur in the present case because $m$ is even. It
follows that $x=-1$ is in all cases a true singularity, which is
null for $-2\le\lambda_+<0$.
\end{enumerate}

\item For $u_0=0$ (again with only the first solution (\ref{J}) for
$J(u)$), the only difference with the preceding analysis is that the
metric (\ref{sinh1}) is non-asymptotically flat (NAF) for $x \to
x_1$. The geodesic equation for the asymptotic metric (\ref{naf})
is:
\begin{equation}\label{gesinhnafi}
\dot{r}^2\sim E^2 -
\bigg(\frac{r}{r_+}\bigg)^{2/\lambda_+}\bigg[\epsilon +
\bigg(\frac{r}{r_+}\bigg)^{2/\lambda_+}\frac{L^2}{r^2}\bigg]\,.
\end{equation}

For $\lambda_+ < 0$, the effective potential goes to zero for $r \to
\infty$, which is at infinite geodesic distance. The conformal
radial coordinate $y$, given asymptotically by $dy
\sim(r/r_+)^{-2/\lambda_+}dr$ also goes to infinity, so that the
conformal metric is asymptotically Minkowskian. The extension across
the horizon $x = 0$ ($r=r_+$) proceeds as in the case $u_0>0$ and
leads to the same Penrose diagrams, i.e. Fig. 1 for $n$ odd and $m$
odd or $m$ even with $\lambda_+<-2$, Fig. 2 for $n$ even and $m$ odd
or $m$ even with $\lambda_+<-2$, Fig. 3 for $n$ odd and $m$ even
with $-2\le\lambda_+<0$, and Fig. 4 for $n$ even and $m$ even with
$-2\le\lambda_+<0$.

For $0<\lambda_+<1$, equation (\ref{gesinhnafi}) leads
asymptotically to $s \sim r^{1-1/\lambda_+}$, so that geodesics
terminate at the conformally timelike singularity $r\to\infty$.
These solutions do not correspond to black holes. However, for
$\lambda_+=1$ ($m=n$), asymptotically $s \sim \ln{r}$, and geodesics
are complete, but spatial infinity is still conformally timelike.
The extension of the metric (\ref{emas}) with $x_1=1$  proceeds
similarly to the case $u>u_0$. For $m=n$ odd, this metric is regular
at $x=-1$ and form-invariant under the combined inversion $x \to z =
1/x$, leading after extension through the two horizons $x=0$ and
$z=0$ to a geodesically complete spacetime (Fig. 9). For $m=n$ even,
the spacetime has a single horizon $x=0$ and a conformally timelike
singularity $x=-1$ (Fig. 10).

\item For $u_0<0$, we have seen that the event horizon at
$u=u_0$ is regular (of order $p$) if  $\lambda_+ = -2/p$
(for any $m$ and $n$ real). The properties of the metric inside the
event horizon ($u < u_0$) depend on the solution (\ref{J}) for
$J(u)$.
\begin{enumerate}
\item First solution (\ref{J}). The form of the metric (\ref{sinh})
shows that $x=0$ ($u\to-\infty$) is a horizon, which is regular if
$m$ and $n$ are positive integers. Because $\lambda_+ = -2/p$, these
must satisfy Eq. (\ref{l2p}). So there are three possibilities:
\begin{enumerate}
\item For $m$ and $n$ generics, $u=-\infty$ is a null singularity.
The Penrose diagram is given by Fig. 3 for $p$ odd, and Fig. 4 for
$p$ even.
\item For $m=n=1$, geodesics terminate at the singularity
$x\rightarrow -\infty$. The Penrose diagram is given by Fig. 6 for
$p$ odd, and by Fig. 8 for $p$ even.
\item For $m=2n+1,p=n-1$, the singularity is again at $x\rightarrow
-\infty$. The Penrose diagram is now given by Fig. 7 for $p$ odd,
and by Fig. 8 for $p$ even.
\end{enumerate}
\item Second solution (\ref{J}). The metric is (\ref{horu0}) with
$b=0$ and $\varphi_1 =\pm a$ (which follows from
(\ref{constraint})), leading to
\be h(u) = e^{\mp a\lambda u}\sinh
a(u-u_0)\,.
\ee

The  associated geodesic equation is
\be \dot{r}^2 =
E^2 - ch^p [\epsilon - L^2ch^p/r^2]\,,\lb{geu1}
\ee
with $r=-1/u >
0$. Near the singularity $r = 0$ ($u \to - \infty$), $h(r) \simeq
e^{(1\pm\lambda)a/r}/2$, so that the analysis follows closely that
made in Subsect. 4.2. Taking into account $\lambda^2 > 1$, there are
two possibilities:
\begin{enumerate}
\item If $\pm\lambda>1$, $h(r)$ diverges, leading to
a space-like singularity if $p$ is odd (Fig. 1), or a time-like
singularity if $p$ is even (Fig. 2).
\item If $\pm\lambda<-1$, $h(r)$ vanishes, signalling a horizon.
However the effective potential in (\ref{geu1}) diverges for $r\to
-0$ ($h(r) \simeq e^{(1\mp\lambda)a/|r|}/2$), so $r=0$ is actually a
null singularity. The Penrose diagram is given by Fig. 3 if $p$ is
odd, and Fig. 4 if $p$ is even.
\end{enumerate}
\item Third solution (\ref{J}). In (\ref{horu0}), $e^{2J} =
\bar{b}^2/\sin^2\bar{b}u$. So the metric has apparent singularities
at $u = u_{s_k} \equiv k\pi/\bar{b}$ ($k$ integer). Putting $u =
u_{s_k} + z$, the metric near $z=0$ goes to the asymptotically flat
form \be dS^2 = c'dt^2 - c^{'-1}(z^{-4}dz^2 + z^{-2}d\Omega^2)\,,
\ee with $c' = ch^p(u_s)$. So, in a generic solution sector $u_{s_k}
< u < u_{s_{k+1}}$ which does not contain $u_0$, we have a
geodesically complete, horizonless spacetime with two asymptotic
regions --- a Lorentzian wormhole \cite{worm} generalizing the
$\lambda = 0$ Bronnikov wormhole of Einstein-Maxwell-anti-scalar
theory \cite{kirill5}. In the solution sector which contains $u_0$,
the spacetime is still geodesically complete with a horizon of order
$p$. The corresponding Penrose diagrams are given in Fig. 11 for $p$
odd, and Fig. 12 for $p$ even.
\end{enumerate}
\end{enumerate}

\subsection{The $a=0$ solutions}
We have seen in Subsect. 3.4 that there are two regular black hole
cases:
\begin{enumerate}
\item For $u_0\geq 0$, $a=b=0$, the metric is given by (\ref{ab0})
with $\lambda_+ = 2/(p+1)$. This has a horizon of order (p+1) at
$r=0$ ($u \to -\infty$), and a singularity at $r = -1/u_0$.
Therefore, for $u_0 > 0$ the Penrose diagram is given by Fig. 1 if
$p$ is even, and Fig. 2 if $p$ is odd.

For $u_0 = 0$, (\ref{ab0}) is replaced by the non-asymptotically
flat metric \be\lb{nafi} dS^2 = \left(\frac{r}{r_+}\right)^{p+1}dt^2
- \left(\frac{r}{r_+}\right)^{-(p+1)}(dr^2 + r^2d\Omega^2)\,. \ee
The two-dimensional reduced metric is similar to that of the
'first-class' black holes of \cite{cold1}. Spacelike infinity $r \to
\infty$ is conformally timelike. For $p$ even, geodesics cross the
odd horizon $r = 0$ and terminate at the conformally spacelike
singularity $r \to -\infty$ (Fig 13). For $p$ odd, $r \to -r$ is an
isometry of the metric (\ref{nafi}). The Penrose diagram of these
geodesically complete spacetimes is given in Fig. 14.

\item For $u_0 < 0$, the metric is (\ref{horu0}) with $h(u)$ given
by (\ref{ha0}). This has a horizon of order $p$ at $u = u_0$ and a
coordinate singularity at $u \to -\infty$. If $b^2 > 0$ ($\varphi_1
\neq 0$), the metric written in terms of the coordinate $x$ contains
again a non-analytic logarithm, so that $u \to -\infty$ ($x=0$) is a
true singularity. If $b^2 = 0$, the metric reduces to (\ref{ab0m}), which
is clearly singular for $u \to -\infty$ ($r = 0$). In both cases, the
Penrose diagram is given by Fig. 1 if $p$ is odd, and Fig. 2 if $p$ is even.

\end{enumerate}

\subsection{The $\sin$ solution}

As seen in subsection (3.5), for this case the only possibility
leading to black holes occurs for the first solution (\ref{J}) for
$J(u)$, and is of the form (\ref{horu0}) with $h(u)$ given by
(\ref{hsin}). This can be rewritten as \be h(u) =
e^{-\lambda\varphi_1u}\sin\bar{a}(u-u_s)\,, \ee for $u_s<u<0$, where
$u_s = u_0 +k\pi/\bar{a}$, with $-\pi/\bar{a}<u_s<0$. The
asymptotically Minkowskian (for $u \to 0$) spacetime therefore
presents an infinite series of regular horizons $u=u_s$, $u=u_s
-\pi/\bar{a}$, $u=u_s-2\pi/\bar{a}$, $\cdots$, all of order $p$,
separating successive regions $I$, $II$, $III$, $\cdots$ which are
all different (because of the non-periodic functions $e^{2J(u)}$ and
$e^{-\lambda\varphi_1u}$). This spacetime is geodesically complete.
For $p$ even, the successive regions all have the same light-cone
orientation. The corresponding Penrose diagram is represented in
Fig. 15. On the other hand, for $p$ odd, the light-cone orientations
alternate between successive regions, so that geodesics can wind
around indefinitely. It is not possible to draw a flat Penrose
diagram for this case.

\begin{figure}[!h]
\centering
\includegraphics[height=2cm,width=3cm]{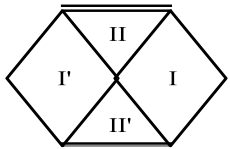}
\caption{\scriptsize{Penrose diagram for the $\cosh$ solution with
$n$ odd and $m$ even ($\lambda_{+}<0$) or $m$ odd ($\lambda_{+}<-2$
or $\lambda_{+}>0$); for the linear solution with $n=m=1$ and $q_1 <
0$; for the $\sinh$ solution with $n$ odd and $m$ odd
($u_0\ge0,\lambda_+<0$) or $m$ even ($u_0\ge0,\lambda_+<-2$ or
$u_0>0,\lambda_+>0$), or with $b=0$ ($u_0<0,\lambda_+ = -2/p,
\pm\lambda > 1$) and $p$ odd; and for the $a=0$ solution with $b=0$
($u_0>0,\lambda_+=2/(p+1)$) and $p$ even or with $b^2\ge0$
($u_0<0,\lambda_+=-2/p$) and $p$ odd. The central singularity is
represented by a double line.}} \label{fig1}
\end{figure}
\begin{figure}[!h]
\centering
\includegraphics[height=3cm,width=3cm]{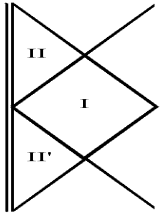}
\caption{\scriptsize{Penrose diagram for the $\cosh$ solution with
$n$ even and $m$ even ($\lambda_{+}<0$) or $m$ odd ($\lambda_{+}<-2$
or $\lambda_{+}>0$); for the $\sinh$ solution with $n$ even and $m$
odd ($u_0\ge0,\lambda_+<0$) or $m$ even ($u_0\ge0,\lambda_+<-2$ or
$u_0>0,\lambda_+>0$), or with $b=0$ ($u_0<0,\lambda_+ = -2/p,
\pm\lambda > 1$) and $p$ even; and for the $a=0$ solution with $b=0$
($u_0>0,\lambda_+=2/(p+1)$) and $p$ odd or with $b^2\ge0$
($u_0<0,\lambda_+=-2/p$) and $p$ even.}} \label{fig2}
\end{figure}
\begin{figure}[h]
\centering
\includegraphics[height=3cm,width=3cm]{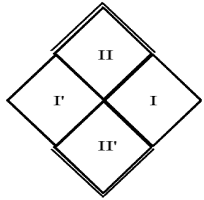}
\caption{\scriptsize{Penrose diagram for the cosh solution with $n$
odd and $m$ even ($\lambda_{+}>0$) or $m$ odd
($-2\leq\lambda_{+}<0$); and for the $\sinh$ solution with $n$ odd
and $m$ odd ($u_0>0, \lambda_+>0$) or $m$ even
($u_0\ge0,-2<\lambda_+<0$), or $u_0<0$ and $\lambda_+=-2/p$ with $p$
odd ($b^2>0$ or $b=0$ and $\pm\lambda<-1$).}} \label{fig3}
\end{figure}
\begin{figure}[h]
\centering
\includegraphics[height=3cm,width=3cm]{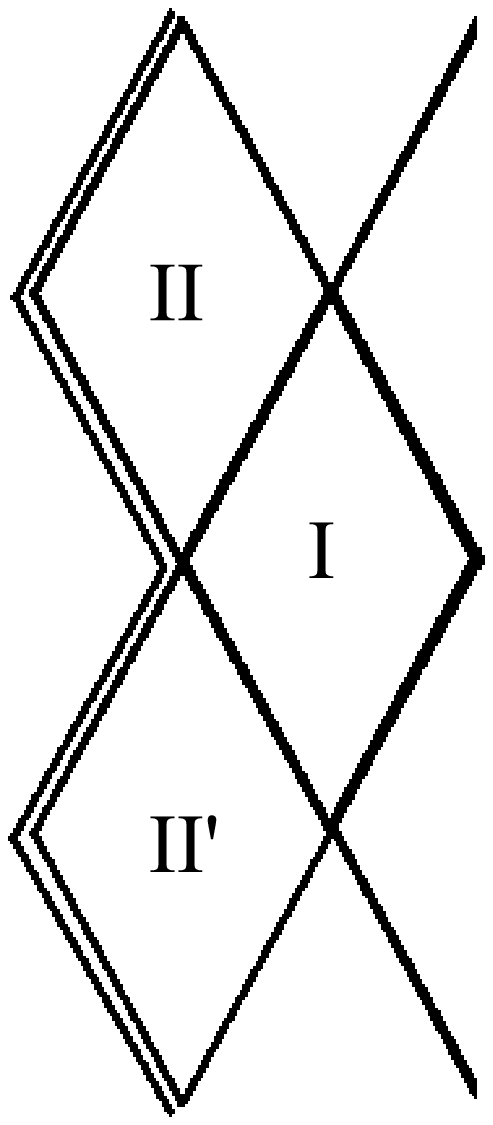}
\caption{\scriptsize{Penrose diagram for the cosh solution with $n$
even and $m$ even ($\lambda_{+}>0$) or $m$ odd
($-2\leq\lambda_{+}<0$); for the linear solution with $n=m=1$ and
$q_1>0$; and for the $\sinh$ solution with $n$ even and $m$ odd
($u_0>0, \lambda_+>0$) or $m$ even ($u_0\ge0,-2<\lambda_+<0$), or
$u_0<0$ and $\lambda_+=-2/p$ with $p$ even ($b^2>0$ or $b=0$ and
$\pm\lambda<-1$).}} \label{fig4}
\end{figure}
\begin{figure}[h]
\centering
\includegraphics[height=5cm,width=5cm]{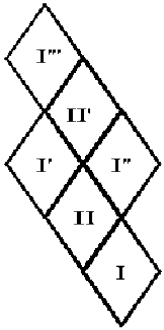}
\caption{\scriptsize{Penrose diagram for the $\cosh$ solution with
$\lambda_+=1$, $m=n$ even}} \label{fig5}
\end{figure}
\begin{figure}[h]
\centering
\includegraphics[height=5cm,width=5cm]{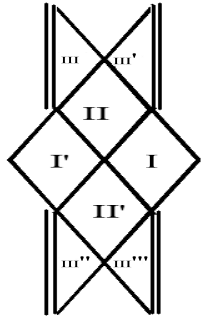}
\caption{\scriptsize{Penrose diagram for the $\cosh$ solution with
$m=n=1$ and $\lambda_+=-2/p$ with $p$ odd; and for the $\sinh$
solution with $u_0>0$ and $\lambda_+=2/(p+1)$ ($m=n=1$) with $p$ odd
or $\lambda_+=2/(n+1)$ ($m=1$) with $n$ odd or $\lambda_+=1$ with
$m=n$ odd, or with $u_0<0$ ($b^2>0,m=n=1$) and $\lambda_+=-2/p$)
with $p$ odd.}} \label{fig6}
\end{figure}
\begin{figure}[h]
\centering
\includegraphics[height=5cm,width=5cm]{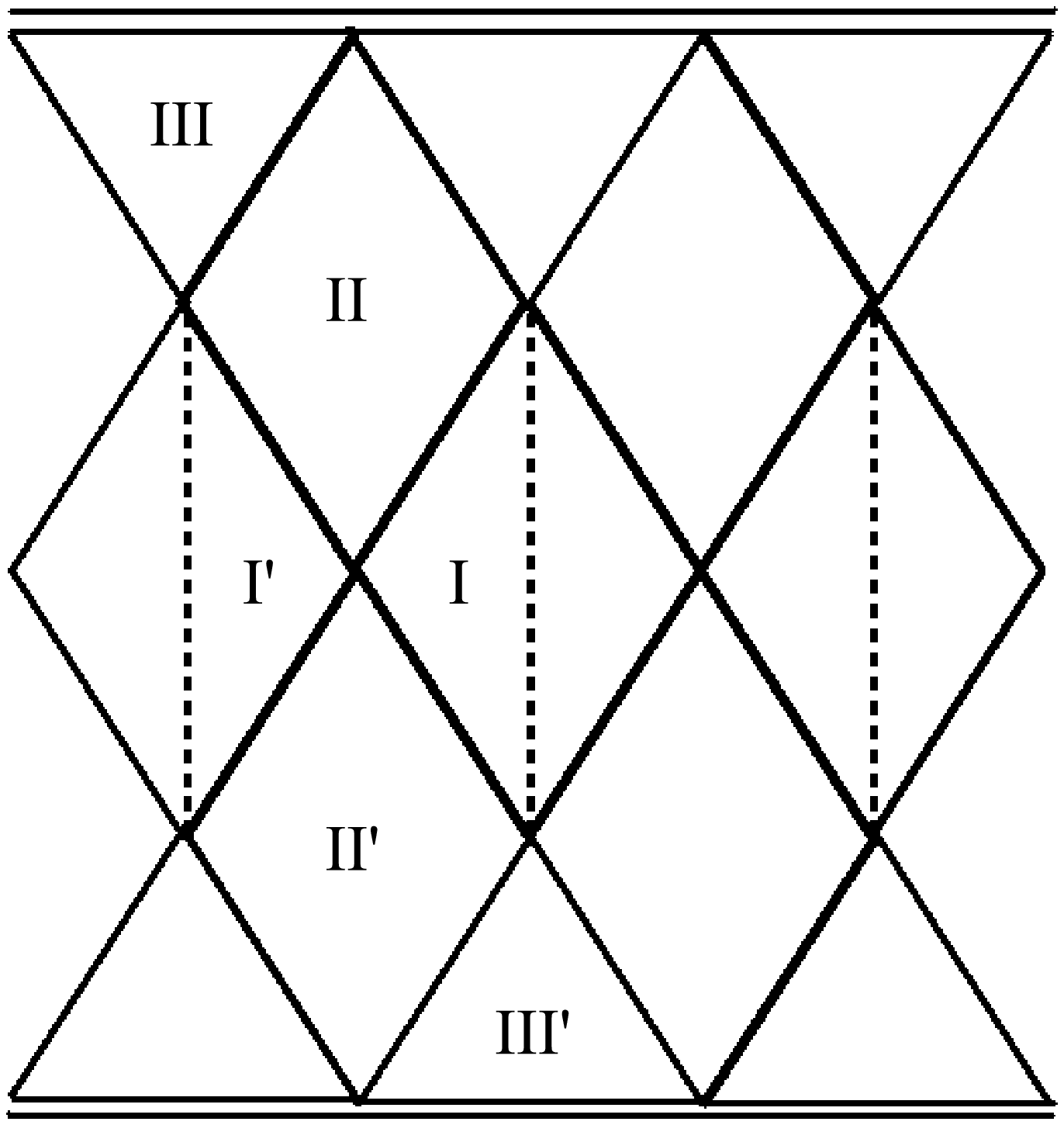}
\caption{\scriptsize{Penrose diagram for the $\cosh$ solution with
$\lambda_+=-2/p$ and $p$ even ($m=n=1$ or $m=2n+1,p=n-1$); and for
the $\sinh$ solution with $u_0>0,\lambda_+=2/(p+1)$ and $p$ even
($m=n=1$), or with $u_0<0,\lambda_+=-2/p$ ($m=2n+1,p=n-1$ odd).}}
\label{fig7}
\end{figure}
\begin{figure}[h]
\centering
\includegraphics[height=5cm,width=5cm]{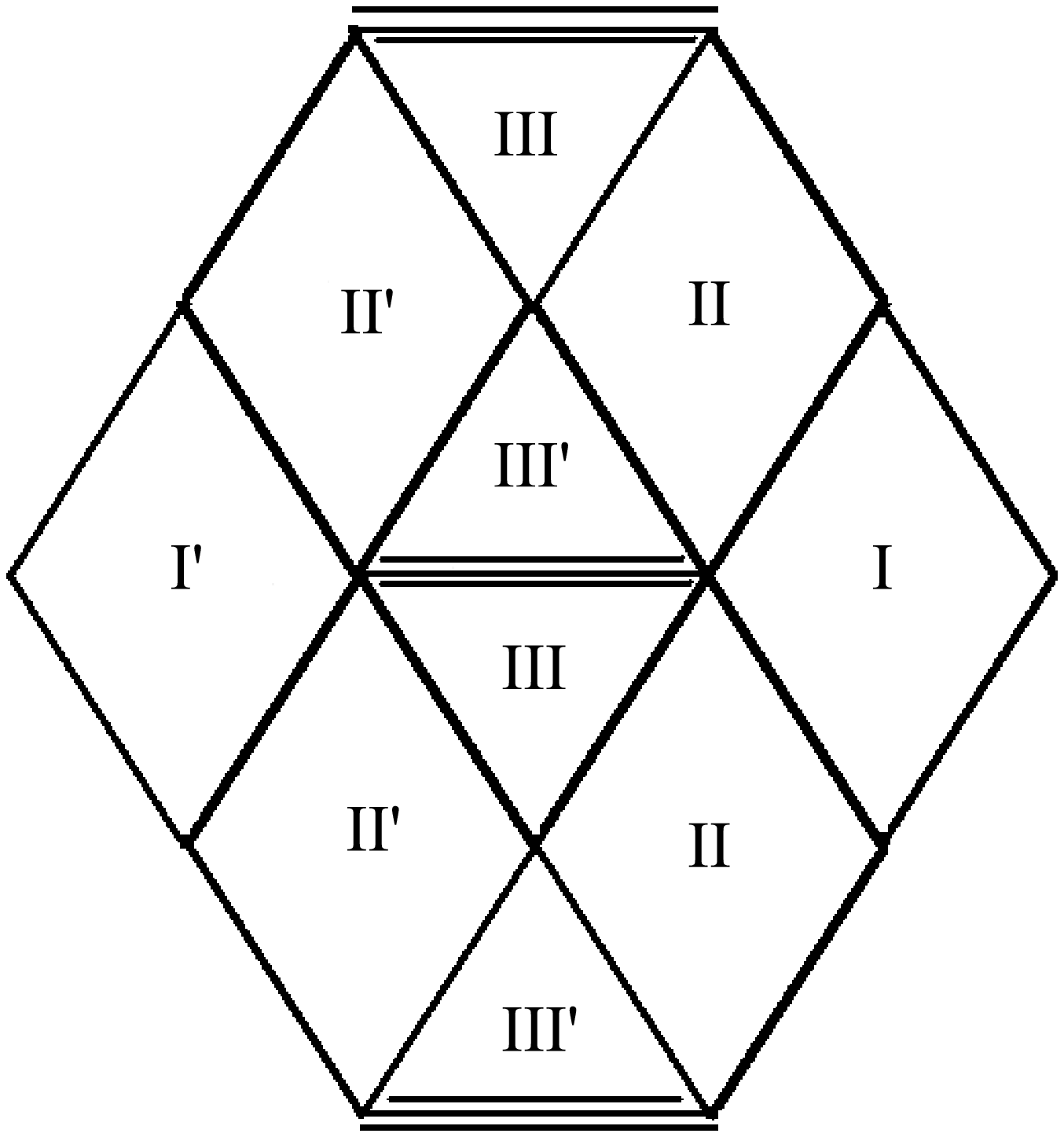}
\caption{\scriptsize{Penrose diagram for the $\cosh$ solution with
$\lambda_+=-2/p$ and $p$ odd ($m=2n+1,p=n-1$); and for the $\sinh$
solution with $u_0>0,\lambda_+=2/(n+1)$ and $n$ even ($m=1$), or
with $u_0<0,\lambda_+=-2/p$ and $p$ even ($m=n=1$ or
$m=2n+1,p=n-1$).}} \label{fig8}
\end{figure}
\begin{figure}[h]
\centering
\includegraphics[height=5cm,width=3cm]{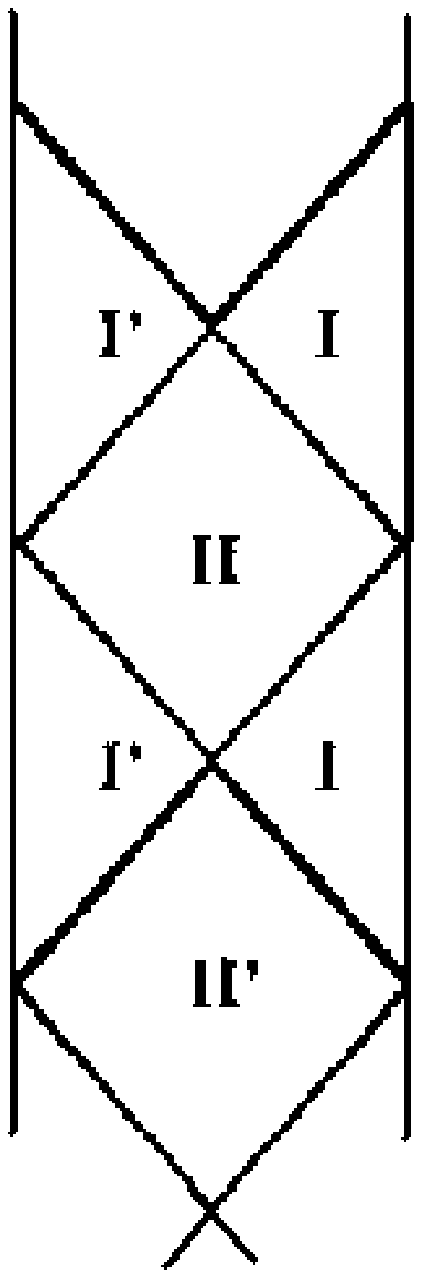}
\caption{\scriptsize{Penrose diagram for the $\sinh$ solution with
$u_0=0,\lambda_+=1,m=n$ odd.}} \label{fig9}
\end{figure}
\begin{figure}[h]
\centering
\includegraphics[height=4cm,width=2cm]{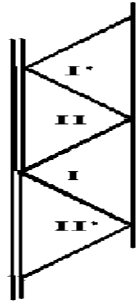}
\caption{\scriptsize{Penrose diagram for the $\sinh$ solution with
$u_0=0,\lambda_+=1,m=n$ even.}} \label{fig10}
\end{figure}
\begin{figure}[h]
\centering
\includegraphics[height=3cm,width=2cm]{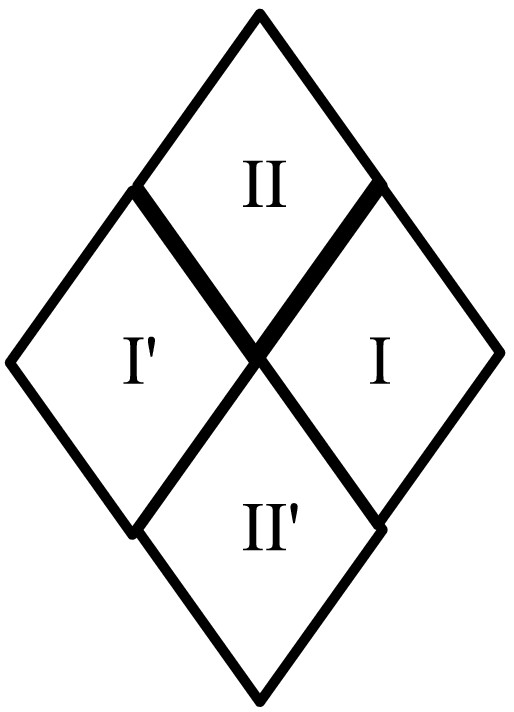}
\caption{\scriptsize{Penrose diagram for the $\sinh$ solution with
$u_0<0$ ($b^2<0$) and $\lambda_+= -2/p$, $p$ odd.}} \label{fig11}
\end{figure}
\begin{figure}[h]
\centering
\includegraphics[height=3cm,width=3cm]{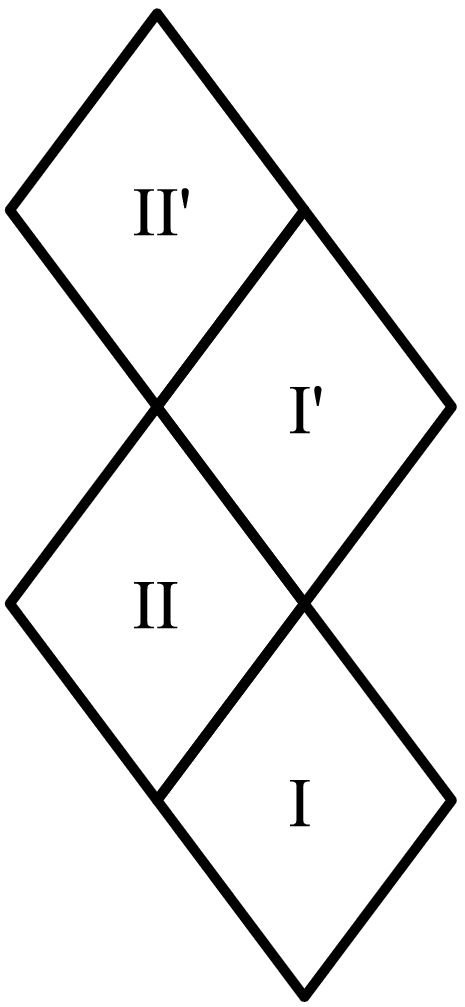}
\caption{\scriptsize{Penrose diagram for the $\sinh$ solution with
$u_0<0$ ($b^2<0$) and $\lambda_+= -2/p$, $p$ even.}} \label{fig12}
\end{figure}
\begin{figure}[h]
\centering
\includegraphics[height=3cm,width=3cm]{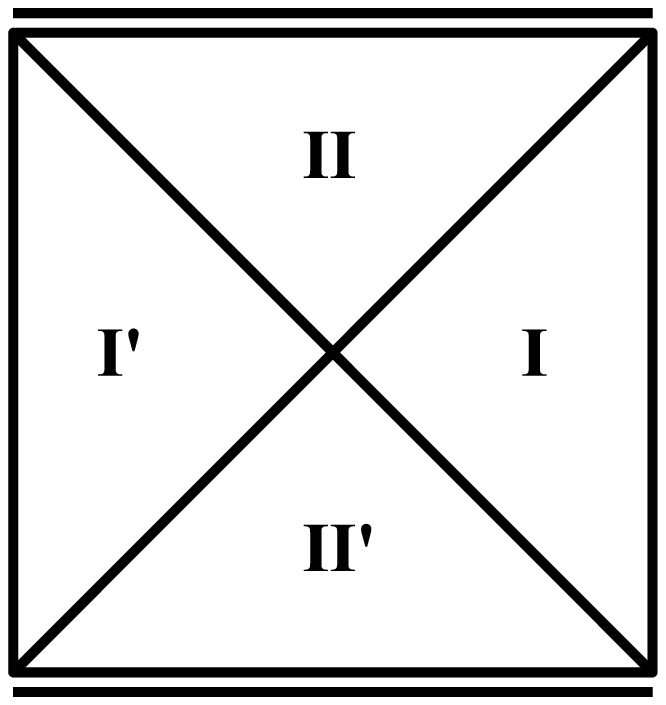}
\caption{\scriptsize{Penrose diagram for the $a=0$ solution
($b=0,u_0=0,\lambda_+= -2/(p+1)$) with $p$ even.}} \label{fig13}
\end{figure}
\begin{figure}[h]
\centering
\includegraphics[height=3cm,width=2cm]{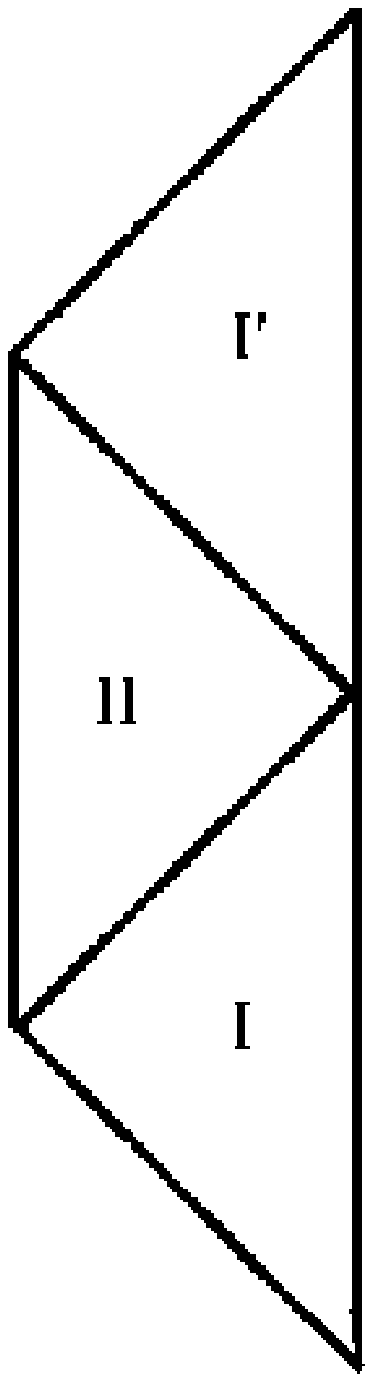}
\caption{\scriptsize{Penrose diagram for the $a=0$ solution
($b=0,u_0=0,\lambda_+= -2/(p+1)$) with $p$ odd.}} \label{fig14}
\end{figure}
\begin{figure}[h]
\centering
\includegraphics[height=5cm,width=3cm]{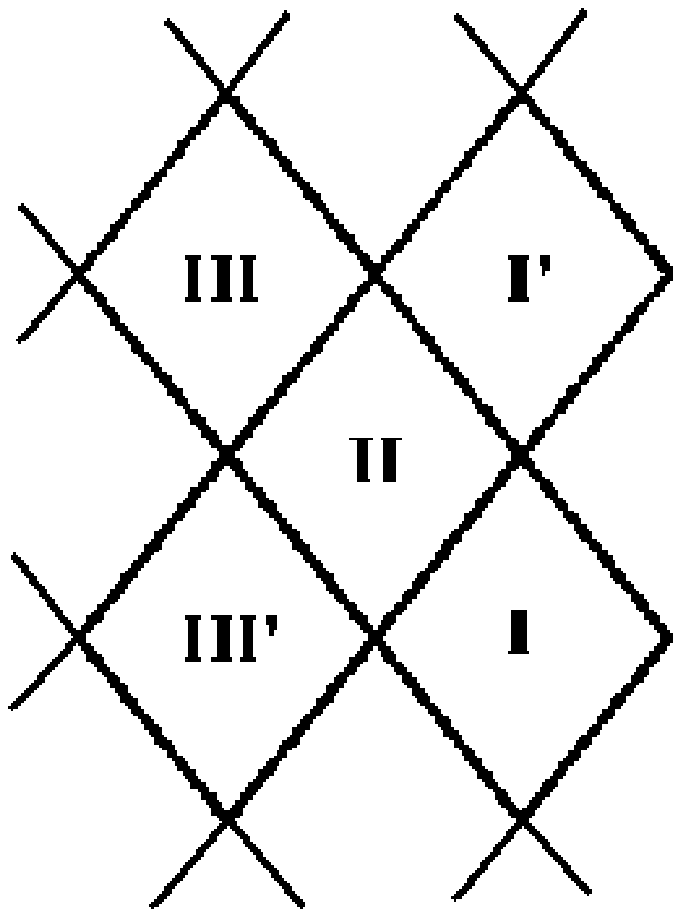}
\caption{\scriptsize{Penrose diagram for the $\sin$ solution with
$\lambda_+= -2/p$, $p$ even.}} \label{fig15}
\end{figure}

\section{Conclusions}

We have determined in this paper the general static, spherically
symmetric solutions for the four-dimensional EMD theory when the
scalar field and/or the electromagnetic field are allowed to violate
the null energy condition. The general solution given by (\ref{om})
contains nine classes of asymptotically flat phantom black holes:
the ``cosh" solution, the ``linear" solution, the ``sinh" solution
with $u_0 > 0$, the ``sinh" solution with $u_0 < 0$ (three classes),
the ``a=0" solution with $u_0 > 0$, the ``a=0" solution with $u_0 <
0$, and the ``sin" solution. There are also two classes of
non-asymptotically flat phantom black holes, corresponding to the
sinh and the a=0 solutions with $u_0=0$.

The event horizon of these black holes can be either non-degenerate
or degenerate. Besides the previously known phantom black holes with
a single event horizon \cite{gibbons,gao}, which occur for generic
values of the dilatonic coupling constant, we have obtained for
certain discrete values of this coupling constant new phantom black
holes with a single event horizon, as well as cold black holes, with
a degenerate event horizon. A noteworthy consequence of the
violation of the null energy condition is that, for the special case
of a vanishing dilaton coupling constant, there is an infinite
sequence of black holes with multiple event horizons, of the cosh
type in the Einstein-anti-Maxwell-anti-scalar case, and of the sinh
type in the Einstein-Maxwell-anti-scalar case.

We have paid special attention to the study of the causal structures
of these phantom black holes. In total, we have found $16$ different
types of causal structures. Many cases lead to Penrose diagrams
similar to those of the ``classical'' Schwarzschild,
Reissner-Nordstr\"om, or extreme Reissner-Nordstr\"om black holes.
However, there are new causal structures. Some of them differ from
the preceding by the fact that the central spacelike or timelike
singularity is replaced by a null singularity (Figs. 3, 4), or that
null infinity is replaced by timelike infinity (Figs. 10, 13). A
number of these spacetimes are geodesically complete with one
degenerate horizon (Figs. 11, 12, 14) or two equally degenerate
horizons (Figs. 5, 9). We also found more complex structures. The
maximal analytic extension of a black hole with an event horizon of
even order and an inner horizon of odd order has a tower of
spacelike singularities (Fig. 8). The opposite case of an odd event
horizon and an even inner horizon cannot be represented by a
two-dimensional conformal diagram, it is possible to represent only
the global spacetime topology (Fig. 7). The most exotic,
geodesically complete, spacetimes have an infinite sequence of
equally degenerate horizons separating successive non-isometrical
regions; this structure is represented in Fig. 15 for even horizons,
the case of oddly-degenerate horizons does not admit a
two-dimensional representation.

There is a discussion in the literature whether phantom black holes
could be formed by gravitational collapse of a phantom fluid
\cite{gao,cai}. In any case, these phantom black holes could perhaps
also be created by another mechanism, such as a tunneling quantum
process. Of course, since for a phantom fluid all energy conditions
are violated, the stability of the configurations described above
remains another important question. The present work should also be
complemented by the construction of rotating phantom black holes,
and the analysis of the solutions corresponding to wormholes, which
we mentioned only briefly at the end of Subsect. 4.3. We hope to
address these questions in the future.

\vspace{0.5cm} \noindent {\bf Acknowledgements:} J.C. Fabris and
M.E.R. thank CNPq (Brazil), FAPES (Brazil) and the French-Brazilian
scientific cooperation CAPES/COFECUB (project number 506/05) for
partial financial support. They also thank the LAPTH
(Annecy-le-Vieux, France), for hospitality during the elaboration of
this work.

\end{document}